\documentclass[aps,prl,twocolumn,superscriptaddress,showpacs]{revtex4} 
\usepackage{graphicx}  
\usepackage{dcolumn}   
\usepackage{bm}        
\usepackage{amssymb}   
\usepackage{amsmath}
\usepackage{verbatim}
\usepackage{color}

\begin{document}

\title{All-optical control of surface plasmons by second-harmonic generation}

\author{Sergio G. Rodrigo}\email{sergut@unizar.es}
\affiliation{Instituto de Ciencia de Materiales de Arag\'{o}n and
Departamento de F\'{i}sica de la Materia Condensada, CSIC-Universidad de Zaragoza, E-50009, Zaragoza, Spain}  
\affiliation{Centro Universitario de la Defensa, Ctra. de Huesca s/n, E-50090 Zaragoza, Spain}

\date{\today}

\begin{abstract}

Light with light control of surface plasmon polaritons is theoretically demonstrated. A barely simple and compact source of these waves consists in a finite number of slits (evenly spaced) perforating a metal film. The system scatters electromagnetic fields in one side of the metal film when it is illuminated from the opposite side by a polarized light source. High intensity light sources moreover efficiently generate light at second harmonic and higher frequencies in the metal led by optical nonlinearities. It is shown how the mixing of fields scattered by the slits from a weak beam at $\lambda$ wavelength, with the second harmonic fields generated by a high intensity $2 \lambda$ beam, creates a destructive interference of surface plasmons in one of the two possible directions of emission from the slits, while these are enhanced along the opposite direction. The unidirectional launching of surface plasmons is due  to the different properties of symmetry at $\lambda$ whether they are linearly or nonlinearly generated. It is envisaged a nanodevice which might allow sending digital information codified in the surface plasmon field or be used to build ultra-narrow bandwidth surface plasmon frequency combs.
\end{abstract}

\maketitle

Short after the first experimental demonstration of a laser device, Franken et al. reported that the high electromagnetic (EM) fields released by lasers allowed the generation of higher frequencies not present in the laser field~\cite{FrankenPRL61}. The first nonlinear harmonic discovered was Second Harmonic Generation (SHG), a process through which two photons with the same frequency generate a new photon with twice the initial frequency. Harmonic generation is behind many of light technologies developed in the past for both research and commercial purposes~\cite{GarmireOptExp13}. Currently new functionalities and applications are being realized for optical characterization and manipulation, sensing and control of light, e.g. based on multipolar nonlinear nanoantennas~\cite{ButetNanoLett12,HaykBookChap12,SmirnovaOptica16} and nonlinear optical metasurfaces~\cite{TymchenkoNature14,MinovichLaser15,WangACSPhot17} which combine metals and different linear and nonlinear materials.

\begin{figure}[thb!]
\centering\includegraphics[width=\columnwidth]{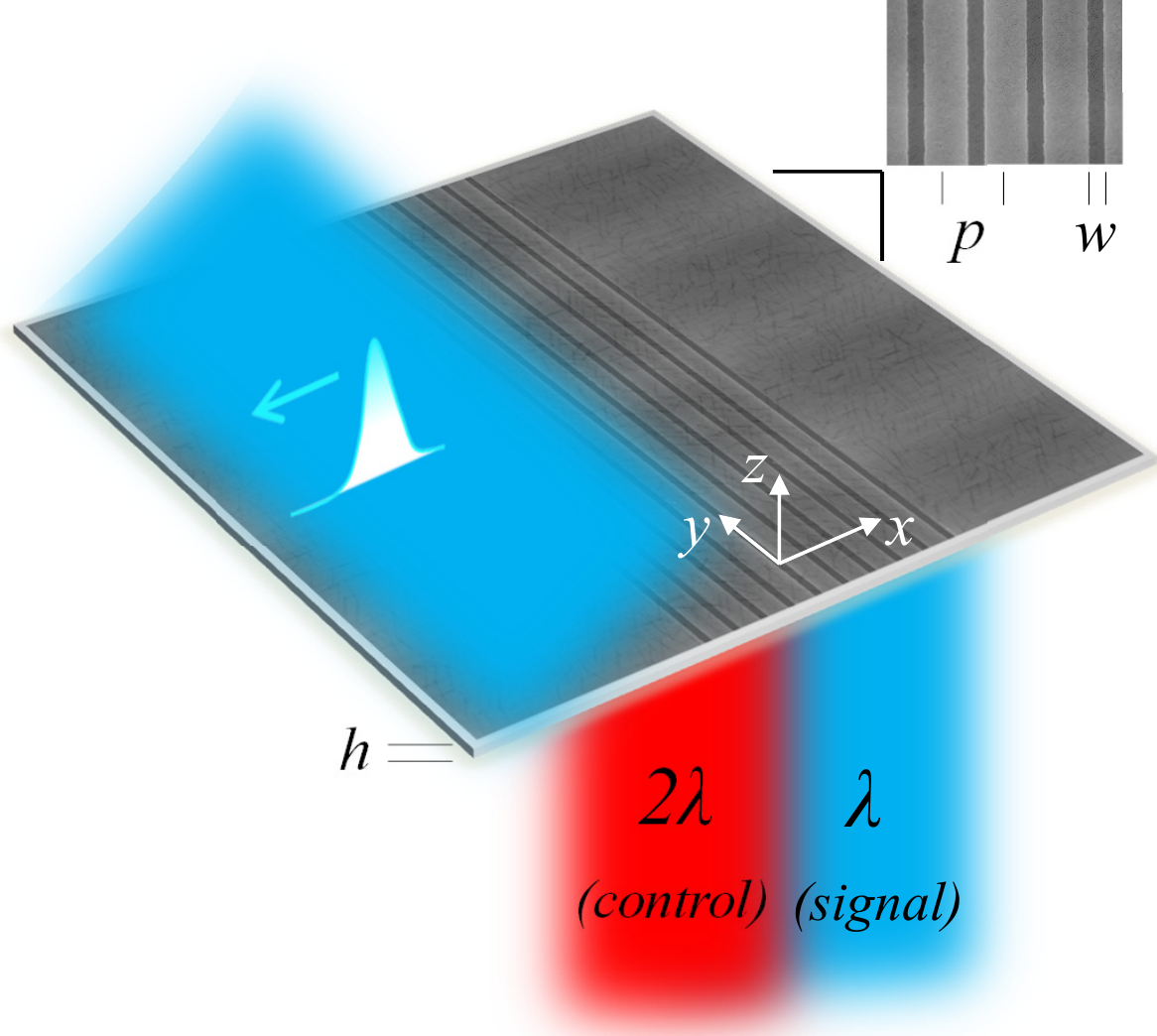} \caption{All-optical control of SPPs by SHG. The simultaneously occurring suppression and enhancement of SPP currents in opposite directions, is a consequence of the different symmetry of SPPs generated by a signal beam at $\lambda$ and the SH field generated by a $2 \lambda$ control beam.} \label{fig1}
\end{figure}

\begin{figure*}
\centering\includegraphics[width=\textwidth]{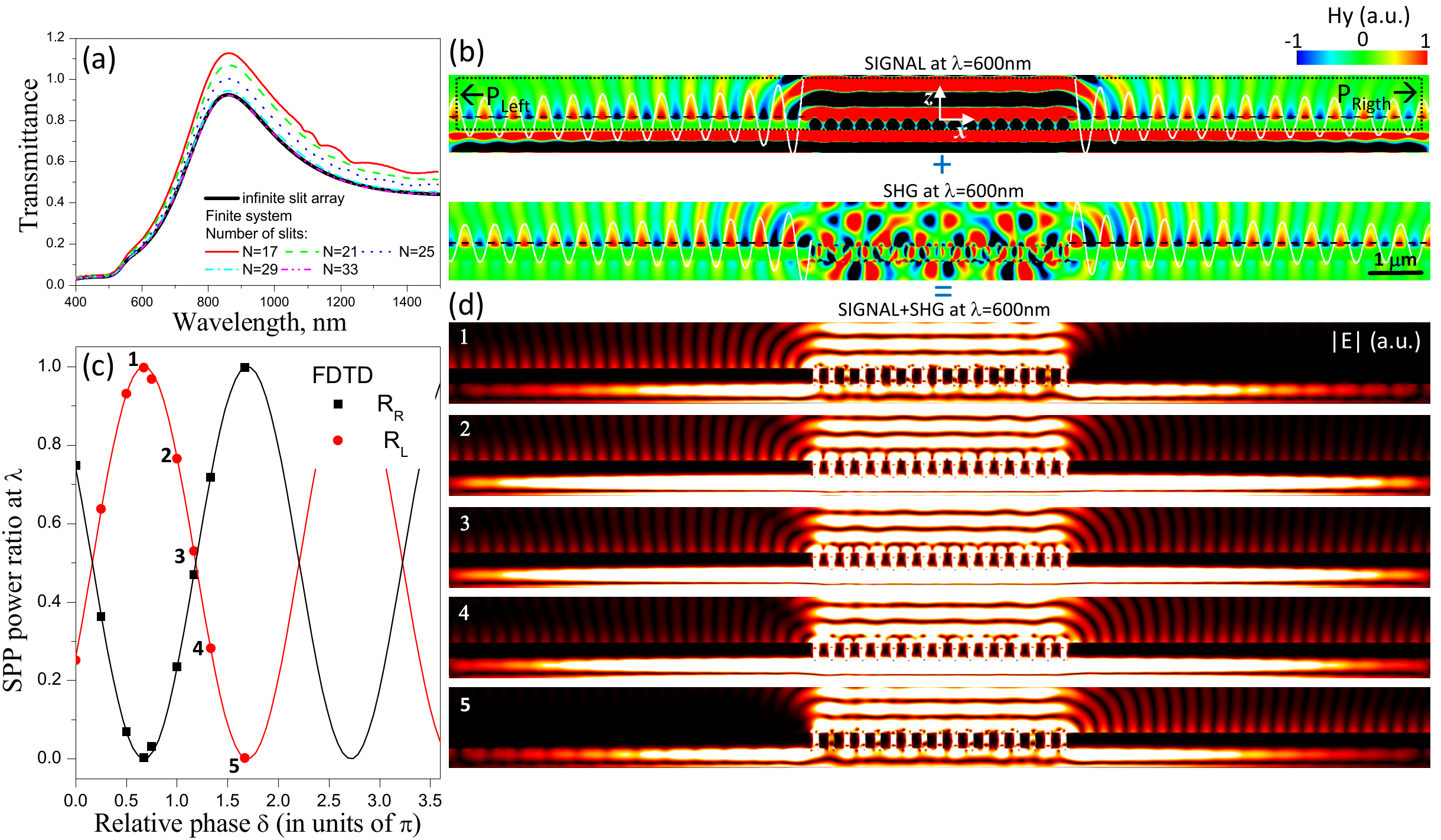} \caption{(a) Linear transmittance through an infinite array of slits (thick line) and different finite systems. In linear and nonlinear calculations the system is illuminated at normal incidence, the electric field  polarized along the x-axis. The geometrical parameters are $p=300$~nm, $w=100$~nm and $h=300$~nm. The system is surrounded by air. (b) Top: real part of the magnetic field component $H_y$ scattered from the signal beam at $\lambda=600$~nm. Bottom: $H_y$ at SH wavelength generated by a $2\lambda$ control beam. The white lines show $H_y$ at $z=0$. (c) Coherent control of SPP directionality by the relative phase $\delta$ between signal and control beams at $\lambda=600$~nm.  The power ratio to the right (black) and to the left (red) are calculated with FDTD as a function of $\delta$ (symbols). The whole behavior is fitted to a sinusoidal function (solid lines). (d) Snapshots of the electric field amplitude for the values of $\delta$ labeled with integer numbers in (c), both beams are switched on. Map scale: red (maximum) and black (minimum).}
\label{fig2}
\end{figure*}

In this Letter, we present a method for actively controlling Surface Plasmon Polaritons (SPP) by means of SHG. These EM modes are bound to metal surfaces~\cite{BarnesNature03}, and it is precisely this confinement which makes them beneficial for light applications at the nanoscale~\cite{AntonioNatPhot17}.  We illustrate our approach using a simple structure: a set of slits perforating an opaque metal film. When illuminated from one side by an external light beam SPP waves are scattered in the opposite side of the film, whenever its electric field oscillates normal to the slit faces. The working principle of our method is depicted in Fig~\ref{fig1}: the SPP fields from two different sources are coherently interfering on the metal surface at the same wavelength $\lambda$; the EM field scattered from a signal beam (central wavelength $\lambda$) by the slits and the one generated by a control beam (central wavelength $2 \lambda$) at its second harmonic (SH). We will show that the EM fields generated at $\lambda$ have different properties of symmetry whether they are linearly generated by the signal beam or originate from SHG. The singular contrast between them will allow us to demonstrate unidirectional emission of SPPs actively controlled by SHG, without surface structuring~\cite{LopezTejeiraNatPhys07} and the possibility to create SPP wavepackets,i.e. SPP \textit{bits}, for sending digital information at the nanoscale or building ultra-narrow bandwidth SPP frequency combs.

The induced polarization currents and optical near fields are calculated using our \textit{homemade} Finite-Difference Time-Domain (FDTD) code~\cite{rodrigoTESIS}. The particular details of the nonlinear numerical treatment can be found  elsewhere~\cite{RodrigoJOSAB15} [see also the Supplemental Material (SM)~\cite{aux}]. Briefly, to calculate the nonlinear response at the SH frequency a perturbative approach is followed, assuming the intensity of the fundamental harmonic (FH) field is not affected by the SH fields. The second-order polarization $\textbf{P}^{(\text{SH})}$ is introduced as a surface-like contribution in the FDTD algorithm ~\cite{GuyotSionnestPRB88,CiraciPRB12}:
\begin{eqnarray}
\textbf{P}_n^{(\text{SH})} &=& \left[\chi^{(2)}_{nnn}
\vert E_n^{(\text{FH})}\vert^2+\chi^{(2)}_{ntt}
\vert E_t^{(\text{FH})}\vert^2 \right] \textbf{n} \nonumber \\
\textbf{P}_t^{(\text{SH})}&=& 2\chi^{(2)}_{tnt}
E_n^{(\text{FH})}\textbf{E}_t^{(\text{FH})} \label{eq1}
\label{Eq1}
\end{eqnarray}
where $n$ and $t$ stand for normal and tangential to the metal surface respectively, and $\chi_{ijk}$ are the non-vanishing components of an effective second-order susceptibility tensor. The FH electric field is taken at the metal surface, and from it $\textbf{P}^{(\text{SH})}$ is calculated at the same location.  We have taken the linear dielectric constant of gold and its $\chi^{(2)}$ from Ref.~\cite{HaoChemPhysLett07} and Ref.~\cite{XiangPRB09}, respectively. We have checked our FDTD program by comparing numerical against analytical results of SHG efficiency in metal surfaces, overall finding good agreement between methods (see SM~\cite{aux}).

In the FDTD simulations the system is illuminated by a low-intensity (real valued) plane wave beam at $\lambda=600$nm (the signal) and by a second high-intensity plane wave beam at $2\lambda$ (the control), as illustrated in  Fig.~\ref{fig1}. The experimental implications of such setup will be discussed later.  We are interested in the SPP field at $\lambda$, which final distribution depends on the local field generated at SH and the near field scattered by the signal beam.  In our proposal only the zero diffraction order is allowed at $\lambda$, which ensures the far-field emission at SH is almost suppressed~\cite{RodrigoJOSAB15}. To do that, we have chosen a finite slit array with period $p=300$~nm. In addition, we take the slit width $w=50$~nm and metal thickness $h=300$~nm, the whole system surrounded by air. The number of slits chosen ($N=17$) are enough to reproduce the main feature observed in the optical transmission spectrum calculated for the infinite structure (see Fig.~\ref{fig2}(a)). Transmission is calculated as the ratio of transmitted power over incident power, the last normalized to the whole area of the slit array. 


Figure~\ref{fig2}(b) (top panel) shows a snapshot of the only non-zero component of the magnetic field allowed by symmetry, $H_y$, when the array is solely illuminated by the signal. It has reflection symmetry regarding the central slit ($x=0$) given that $H_y(-\vert x \vert,z)=H_y(\vert x \vert,z)$. The field pattern in air demonstrates we succeed in designing the structure because only the zero diffraction order is seen in the scattered field of the signal beam. In Fig.~\ref{fig2}(b) (bottom panel) the system is illuminated by the control beam (signal \textit{off}). The magnetic field at its SH present point symmetry, that is $H_y(-\vert x \vert,z)=-H_y(\vert x \vert,z)$. The EM fields at the metal surface ``inherit'' the symmetry of the EM waveguide modes excited inside the slits, which is different in each case~\cite{RodrigoJOSAB15}. It is worth to mention that the waveguide modes at SHG are unaccessible in the linear regime at normal incidence. 

Therefore, by switching on signal and control beams simultaneously the final EM field has not definite symmetry because it results from a sum of an odd function plus an even function. The interference between the two terms can be tuned by adjusting the relative intensity and/or phase between them, thus $ \textbf{E}^{2\lambda}_{control}=\alpha \textbf{E}^{\lambda}_{signal}$, where $\alpha = \alpha_0
\exp{(i\delta\pi)}$,  and the relative electric field amplitude $\alpha_0$ is a function of the incident wavelength (depending on the linear and nonlinear optical properties of the materials) and the geometry of the system. To characterize the optical response in terms of directionality of SPP emission we define the SPP power ratio, for instance, to the right as: $R_{R}=P_{R}/(P_{R}+P_{L})$, where $P_{R}$ and $P_{L}$ are the SPP powers to the right and left and calculated on the vertical surfaces represented with dashed lines in Fig.~\ref{fig2}(b) top panel. These surfaces are located far from the slit array and the area of integration has subwavelength transverse size, so the contribution of radiation from the slits to $P_{R}$ and $P_{L}$ is negligible. 

The array of slits is illuminated by both signal and control beams and the results are shown in Fig.~\ref{fig2}(c)-(d). For the structure investigated $\alpha_0 \sim 2200$. The main result of this Letter is that there exist phase conditions for total suppression of SPPs in one direction from the slits, while they are simultaneously enhanced along the opposite direction. In Fig.~\ref{fig2}(c) the SPP power ratios to the right and left are shown for different values of the relative phase $\delta$, calculated with FDTD (symbols). The trend is that of a simple sinusoidal function demonstrating the possibility of continuous coherent control of SPPs by SHG (solid lines). This fact is made more evident from the snapshots of the electric field amplitude shown in Fig.~\ref{fig2}(d) we calculated for different $\delta$ values (those labeled with integer numbers in Fig.~\ref{fig2}(c)). Top and bottom panels visualize the unidirectional launching of SPPs actively controlled by SHG.

\begin{figure}[!t]
\centering\includegraphics[width=\columnwidth]{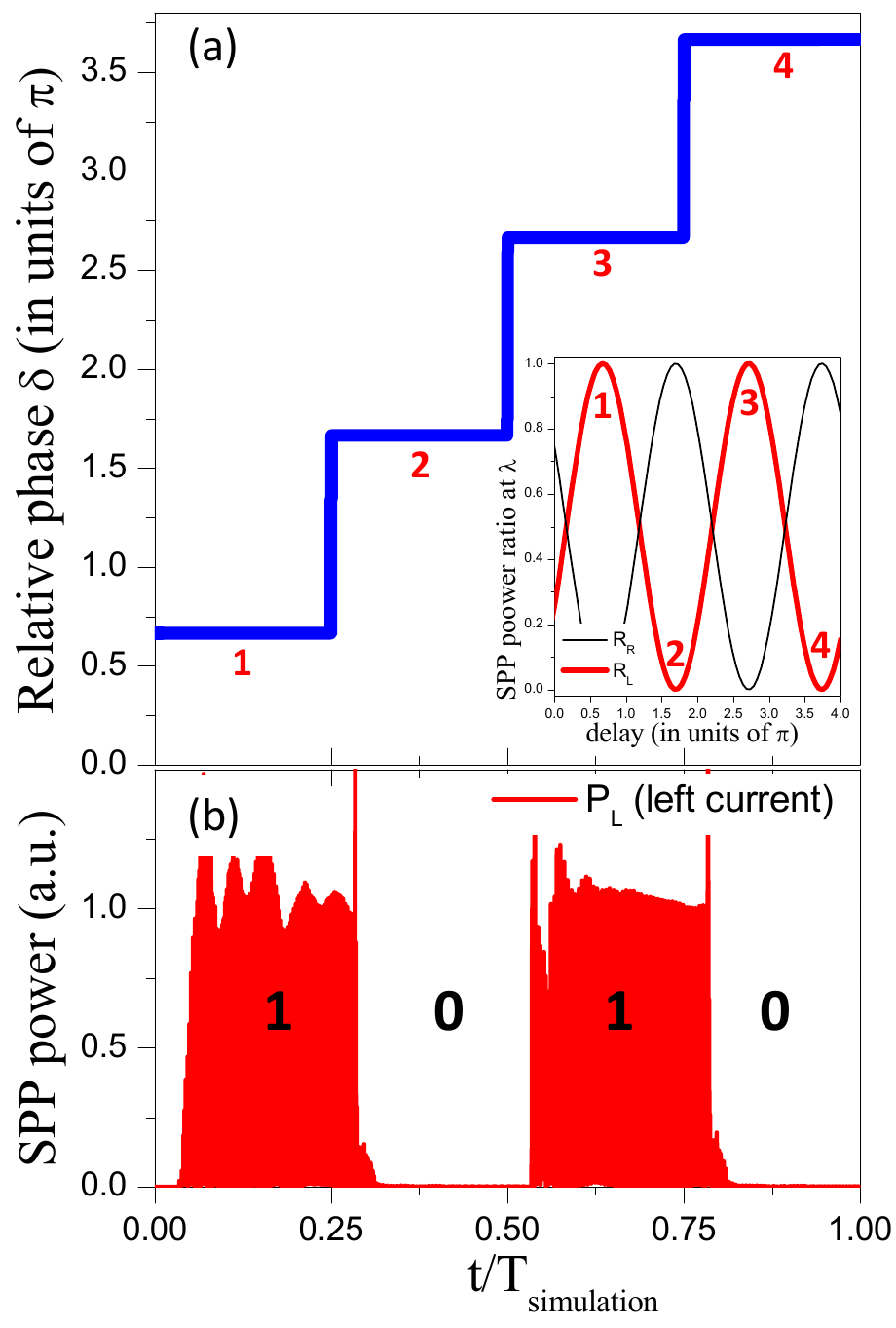}
\caption{Demonstration of dynamical control of SPPs  at $\lambda=600$~nm. During a single FDTD simulation spanning $T_{\text{simulation}} \sim 1$~ps in time, the relative phase between signal and control beams is periodically changed as shown in (a). These values correspond to those labeled with integer numbers in the inset, which corresponds to an enhancement-suppression sequence of the SPP current $P_{L}$. (b) $P_{L}$ as a function of time consists in a train of pulses (\textit{bits}). }
\label{fig3}
\end{figure}

To observe the coherent effect, contributions from signal and SH fields generated by the control beam must be of similar magnitude (adjusted by tuning $\alpha_0$). In addition, there is another constraint: $\textbf{E}^{\lambda}_{signal}$ and $\textbf{E}^{2\lambda}_{control}$ fields have different frequencies and their phases need to be locked, like in experiments for Terahertz wave generation~\cite{LiuNatPhot11,DaiIEEE11}. In these experiments the control beam is focused onto a SHG crystal to generate enough output with a fixed phase relation between signal and control pulses.

In the following we discuss the possibility to dynamically change the relative phase between signal and control beams to achieve a train of SPP pulses (\textit{bits}). Such a device might allow sending digital information codified in the surface plasmon field or serve to produce SPP frequency combs. As a proof of principle, we demonstrate dynamical control of SPPs in Fig.~\ref{fig3}. A single FDTD simulation is set, spanning $T_{\text{simulation}}\sim 1$~ps in time. This short time has been chosen to reduce computational burden. The relative phase between signal and control beams is periodically changed every $T_{\text{simulation}}/4$ time steps as shown in Fig.~\ref{fig3}(a), for values of $\delta$ labeled with integer numbers (see inset). The evenly spaced in time phase dynamics results in a suppression - enhancement sequence in time of the SPP current at the surface. In Fig.~\ref{fig3}(b) the SPP power $P_{L}$ at $\lambda$ is calculated, showing a train of SPP \textit{bits} along the left path equally spaced in time, as expected. 

The bit rate is approximately 1~Gbit/s in the simulations, a number which is of course constrained by our ability to externally tune the time span necessary to switch the phase or the relative intensity from \textit{off} to \textit{on} states and to measure it, with current lasers  able to deliver fempto-second pulses with repetition rates as high as 10~GHz~\cite{BartelsScience09}.


The use of SHG for coherently controlling radiation has been recently purposed in isolated metallic wires~\cite{RodrigoPRL13}. Several drawbacks are inherent to the study of SHG from single nanoparticles, but even for flat metal surfaces there is still some controversy given the subtle nature of SHG processes in metals~\cite{GrossePRL12,CiraciPRB12,NireekshanJOSAB17}. Even though most of the  basic research on SHG from metals was initially focused on metal nanoparticles~\cite{AgarwalSolidStateCommun82,HuaPRB86,OestlingZPhysD93,DadapPRL99,CiraciPRB12,HaykBookChap12} (to name a few), note that only recently the optical characterization of SHG emission from a single nanoparticle has been experimentally atttained~\cite{ButetNanoLett10}. Complex setups are needed to investigate far-field SHG from single metallic nanoparticles as compared to extended systems like arrays of symmetric nanoparticles~\cite{McMahonPRB06} and holey metal films ~\cite{NieuwstadtPRL06}. 

The coherent control of SPPs is not restricted to the geometry here investigated and structures including metallic nanoparticles (e.g. patch antennas~\cite{ManjareAPL17})) and holey metal films can be considered. We additionally provide in the SM~\cite{aux} results for a single metallic nanowire located over an optically thin metal film and we demonstrate coherent control of Long Range-SPPs~\cite{EconomouPhysRev69}. 

Not only active control of confined modes like SPPs or Long-Range SPPs can be attainable using our method, but coherent control of radiation from the corrugated metal surface, as shown in the SM~\cite{aux}.

In conclusion, we have theoretically demonstrated the possibility of building SPP sources actively controlled by an external light source that would allow sending information as SPP \textit{bits} on metal surfaces or be used to produce SPP frequency combs. Our theoretical proposal can be experimentally verified using standard pump-probe spectroscopy, the actual SPP \textit{bit} rate constrained by the relative phase and intensity or phase delay introduced between the signal and control beams that is experimentally achievable. We envisage nonlinear nanodevices for control and generation of light based on coherently combining different light harmonics.  Our proposal is of general application and would work in different plasmonic platforms and frequency regimes.

\begin{acknowledgments}
We acknowledge support from the Spanish Ministry of Economy and Competitiveness under projects: MAT2014-53432-C5-1-R and MAT2017-88358-C3-2-R.
\end{acknowledgments}


\begin{thebibliography}{33}
\expandafter\ifx\csname natexlab\endcsname\relax\def\natexlab#1{#1}\fi
\expandafter\ifx\csname bibnamefont\endcsname\relax
  \def\bibnamefont#1{#1}\fi
\expandafter\ifx\csname bibfnamefont\endcsname\relax
  \def\bibfnamefont#1{#1}\fi
\expandafter\ifx\csname citenamefont\endcsname\relax
  \def\citenamefont#1{#1}\fi
\expandafter\ifx\csname url\endcsname\relax
  \def\url#1{\texttt{#1}}\fi
\expandafter\ifx\csname urlprefix\endcsname\relax\def\urlprefix{URL }\fi
\providecommand{\bibinfo}[2]{#2}
\providecommand{\eprint}[2][]{\url{#2}}

\bibitem[{\citenamefont{Franken et~al.}(1961)\citenamefont{Franken, Hill,
  Peters, and Weinreich}}]{FrankenPRL61}
\bibinfo{author}{\bibfnamefont{P.~A.} \bibnamefont{Franken}},
  \bibinfo{author}{\bibfnamefont{A.~E.} \bibnamefont{Hill}},
  \bibinfo{author}{\bibfnamefont{C.~W.} \bibnamefont{Peters}},
  \bibnamefont{and}
  \bibinfo{author}{\bibfnamefont{G.}~\bibnamefont{Weinreich}},
  \bibinfo{journal}{Phys. Rev. Lett.} \textbf{\bibinfo{volume}{7}},
  \bibinfo{pages}{118} (\bibinfo{year}{1961}).

\bibitem[{\citenamefont{Garmire}(2013)}]{GarmireOptExp13}
\bibinfo{author}{\bibfnamefont{E.}~\bibnamefont{Garmire}},
  \bibinfo{journal}{Opt. Express} \textbf{\bibinfo{volume}{21}},
  \bibinfo{pages}{30532} (\bibinfo{year}{2013}).

\bibitem[{\citenamefont{Butet et~al.}(2012)\citenamefont{Butet,
  Russier-Antoine, Jonin, Lascoux, Benichou, and Brevet}}]{ButetNanoLett12}
\bibinfo{author}{\bibfnamefont{J.}~\bibnamefont{Butet}},
  \bibinfo{author}{\bibfnamefont{I.}~\bibnamefont{Russier-Antoine}},
  \bibinfo{author}{\bibfnamefont{C.}~\bibnamefont{Jonin}},
  \bibinfo{author}{\bibfnamefont{N.}~\bibnamefont{Lascoux}},
  \bibinfo{author}{\bibfnamefont{E.}~\bibnamefont{Benichou}}, \bibnamefont{and}
  \bibinfo{author}{\bibfnamefont{P.-F.} \bibnamefont{Brevet}},
  \bibinfo{journal}{Nano Letters} \textbf{\bibinfo{volume}{12}},
  \bibinfo{pages}{1697} (\bibinfo{year}{2012}).

\bibitem[{\citenamefont{Harutyunyan et~al.}(2012)\citenamefont{Harutyunyan,
  Volpe, and Novotny}}]{HaykBookChap12}
\bibinfo{author}{\bibfnamefont{H.}~\bibnamefont{Harutyunyan}},
  \bibinfo{author}{\bibfnamefont{G.}~\bibnamefont{Volpe}}, \bibnamefont{and}
  \bibinfo{author}{\bibfnamefont{L.}~\bibnamefont{Novotny}},
  \emph{\bibinfo{title}{{Optical Antennas}}} (\bibinfo{publisher}{Cambridge
  University Press}, \bibinfo{address}{Cambridge}, \bibinfo{year}{2012}), chap.
  \bibinfo{chapter}{Nonlinear optical antennas}.

\bibitem[{\citenamefont{Smirnova and Kivshar}(2016)}]{SmirnovaOptica16}
\bibinfo{author}{\bibfnamefont{D.}~\bibnamefont{Smirnova}} \bibnamefont{and}
  \bibinfo{author}{\bibfnamefont{Y.~S.} \bibnamefont{Kivshar}},
  \bibinfo{journal}{Optica} \textbf{\bibinfo{volume}{3}}, \bibinfo{pages}{1241}
  (\bibinfo{year}{2016}).

\bibitem[{\citenamefont{Lee et~al.}(2014)\citenamefont{Lee, Tymchenko,
  Argyropoulos, Chen, Lu, Demmerle, Boehm, Amann, Al\'u, and
  Belkin}}]{TymchenkoNature14}
\bibinfo{author}{\bibfnamefont{J.}~\bibnamefont{Lee}},
  \bibinfo{author}{\bibfnamefont{M.}~\bibnamefont{Tymchenko}},
  \bibinfo{author}{\bibfnamefont{C.}~\bibnamefont{Argyropoulos}},
  \bibinfo{author}{\bibfnamefont{P.-Y.} \bibnamefont{Chen}},
  \bibinfo{author}{\bibfnamefont{F.}~\bibnamefont{Lu}},
  \bibinfo{author}{\bibfnamefont{F.}~\bibnamefont{Demmerle}},
  \bibinfo{author}{\bibfnamefont{G.}~\bibnamefont{Boehm}},
  \bibinfo{author}{\bibfnamefont{M.-C.} \bibnamefont{Amann}},
  \bibinfo{author}{\bibfnamefont{A.}~\bibnamefont{Al\'u}}, \bibnamefont{and}
  \bibinfo{author}{\bibfnamefont{M.~A.} \bibnamefont{Belkin}},
  \bibinfo{journal}{Nature} \textbf{\bibinfo{volume}{511}}, \bibinfo{pages}{65}
  (\bibinfo{year}{2014}).

\bibitem[{\citenamefont{Minovich et~al.}(2015)\citenamefont{Minovich,
  Miroshnichenko, Bykov, Murzina, Neshev, and Kivshar}}]{MinovichLaser15}
\bibinfo{author}{\bibfnamefont{A.~E.} \bibnamefont{Minovich}},
  \bibinfo{author}{\bibfnamefont{A.~E.} \bibnamefont{Miroshnichenko}},
  \bibinfo{author}{\bibfnamefont{A.~Y.} \bibnamefont{Bykov}},
  \bibinfo{author}{\bibfnamefont{T.~V.} \bibnamefont{Murzina}},
  \bibinfo{author}{\bibfnamefont{D.~N.} \bibnamefont{Neshev}},
  \bibnamefont{and} \bibinfo{author}{\bibfnamefont{Y.~S.}
  \bibnamefont{Kivshar}}, \bibinfo{journal}{Laser and Photonics Reviews}
  \textbf{\bibinfo{volume}{9}} (\bibinfo{year}{2015}).

\bibitem[{\citenamefont{Wang et~al.}(2017)\citenamefont{Wang, Martinson, and
  Harutyunyan}}]{WangACSPhot17}
\bibinfo{author}{\bibfnamefont{F.}~\bibnamefont{Wang}},
  \bibinfo{author}{\bibfnamefont{A.~B.~F.} \bibnamefont{Martinson}},
  \bibnamefont{and}
  \bibinfo{author}{\bibfnamefont{H.}~\bibnamefont{Harutyunyan}},
  \bibinfo{journal}{ACS Photonics} \textbf{\bibinfo{volume}{4}},
  \bibinfo{pages}{1188} (\bibinfo{year}{2017}).

\bibitem[{\citenamefont{Barnes et~al.}(2003)\citenamefont{Barnes, Dereux, and
  Ebbesen}}]{BarnesNature03}
\bibinfo{author}{\bibfnamefont{W.}~\bibnamefont{Barnes}},
  \bibinfo{author}{\bibfnamefont{A.}~\bibnamefont{Dereux}}, \bibnamefont{and}
  \bibinfo{author}{\bibfnamefont{T.}~\bibnamefont{Ebbesen}},
  \bibinfo{journal}{Nature} \textbf{\bibinfo{volume}{424}},
  \bibinfo{pages}{824} (\bibinfo{year}{2003}).

\bibitem[{\citenamefont{Fern\'andez-Dom\'inguez
  et~al.}(2017)\citenamefont{Fern\'andez-Dom\'inguez, Garc\'ia-Vidal, and
  Mart\'in-Moreno}}]{AntonioNatPhot17}
\bibinfo{author}{\bibfnamefont{A.~I.} \bibnamefont{Fern\'andez-Dom\'inguez}},
  \bibinfo{author}{\bibfnamefont{F.~J.} \bibnamefont{Garc\'ia-Vidal}},
  \bibnamefont{and}
  \bibinfo{author}{\bibfnamefont{L.}~\bibnamefont{Mart\'in-Moreno}},
  \bibinfo{journal}{Nature Photonics} \textbf{\bibinfo{volume}{11}},
  \bibinfo{pages}{8} (\bibinfo{year}{2017}).

\bibitem[{\citenamefont{Lopez-Tejeira et~al.}(2007)\citenamefont{Lopez-Tejeira,
  Rodrigo, Martin-Moreno, Garcia-Vidal, Devaux, Ebbesen, Krenn, Radko,
  Bozhevolnyi, Gonzalez et~al.}}]{LopezTejeiraNatPhys07}
\bibinfo{author}{\bibfnamefont{F.}~\bibnamefont{Lopez-Tejeira}},
  \bibinfo{author}{\bibfnamefont{S.~G.} \bibnamefont{Rodrigo}},
  \bibinfo{author}{\bibfnamefont{L.}~\bibnamefont{Martin-Moreno}},
  \bibinfo{author}{\bibfnamefont{F.~J.} \bibnamefont{Garcia-Vidal}},
  \bibinfo{author}{\bibfnamefont{E.}~\bibnamefont{Devaux}},
  \bibinfo{author}{\bibfnamefont{T.~W.} \bibnamefont{Ebbesen}},
  \bibinfo{author}{\bibfnamefont{J.~R.} \bibnamefont{Krenn}},
  \bibinfo{author}{\bibfnamefont{I.~P.} \bibnamefont{Radko}},
  \bibinfo{author}{\bibfnamefont{S.~I.} \bibnamefont{Bozhevolnyi}},
  \bibinfo{author}{\bibfnamefont{M.~U.} \bibnamefont{Gonzalez}},
  \bibnamefont{et~al.}, \bibinfo{journal}{Nature Phys.}
  \textbf{\bibinfo{volume}{3}}, \bibinfo{pages}{324} (\bibinfo{year}{2007}).

\bibitem[{\citenamefont{Rodrigo}(2011)}]{rodrigoTESIS}
\bibinfo{author}{\bibfnamefont{S.~G.} \bibnamefont{Rodrigo}},
  \emph{\bibinfo{title}{Optical Properties of Nanostructured Metallic Systems:
  Studied with the Finite-Difference Time-Domain Method}}, Springer Theses
  (\bibinfo{publisher}{Springer}, \bibinfo{year}{2011}).

\bibitem[{\citenamefont{Rodrigo et~al.}(2015)\citenamefont{Rodrigo, Laliena,
  and Mart\'{i}n-Moreno}}]{RodrigoJOSAB15}
\bibinfo{author}{\bibfnamefont{S.~G.} \bibnamefont{Rodrigo}},
  \bibinfo{author}{\bibfnamefont{V.}~\bibnamefont{Laliena}}, \bibnamefont{and}
  \bibinfo{author}{\bibfnamefont{L.}~\bibnamefont{Mart\'{i}n-Moreno}},
  \bibinfo{journal}{J. Opt. Soc. Am. B} \textbf{\bibinfo{volume}{32}},
  \bibinfo{pages}{15} (\bibinfo{year}{2015}).

\bibitem[{aux()}]{aux}
\bibinfo{note}{See Supplemental Material.}

\bibitem[{\citenamefont{Guyot-Sionnest and Shen}(1988)}]{GuyotSionnestPRB88}
\bibinfo{author}{\bibfnamefont{P.}~\bibnamefont{Guyot-Sionnest}}
  \bibnamefont{and} \bibinfo{author}{\bibfnamefont{Y.R.}~\bibnamefont{Shen}},
  \bibinfo{journal}{Phys. Rev. B} \textbf{\bibinfo{volume}{38}},
  \bibinfo{pages}{7985} (\bibinfo{year}{1988}).

\bibitem[{\citenamefont{Cirac\`\i et~al.}(2012)\citenamefont{Cirac\`\i,
  Poutrina, Scalora, and Smith}}]{CiraciPRB12}
\bibinfo{author}{\bibfnamefont{C.}~\bibnamefont{Cirac\`\i}},
  \bibinfo{author}{\bibfnamefont{E.}~\bibnamefont{Poutrina}},
  \bibinfo{author}{\bibfnamefont{M.}~\bibnamefont{Scalora}}, \bibnamefont{and}
  \bibinfo{author}{\bibfnamefont{D.~R.} \bibnamefont{Smith}},
  \bibinfo{journal}{Phys. Rev. B} \textbf{\bibinfo{volume}{86}},
  \bibinfo{pages}{115451} (\bibinfo{year}{2012}).

\bibitem[{\citenamefont{Hao and Nordlander}(2007)}]{HaoChemPhysLett07}
\bibinfo{author}{\bibfnamefont{F.}~\bibnamefont{Hao}} \bibnamefont{and}
  \bibinfo{author}{\bibfnamefont{P.}~\bibnamefont{Nordlander}},
  \bibinfo{journal}{Chem. Phys. Lett.} \textbf{\bibinfo{volume}{446}},
  \bibinfo{pages}{115 } (\bibinfo{year}{2007}).

\bibitem[{\citenamefont{Wang et~al.}(2009)\citenamefont{Wang, Rodr\'iguez,
  Albers, Ahorinta, Sipe, and Kauranen}}]{XiangPRB09}
\bibinfo{author}{\bibfnamefont{F.~X.} \bibnamefont{Wang}},
  \bibinfo{author}{\bibfnamefont{F.~J.} \bibnamefont{Rodr\'iguez}},
  \bibinfo{author}{\bibfnamefont{W.~M.} \bibnamefont{Albers}},
  \bibinfo{author}{\bibfnamefont{R.}~\bibnamefont{Ahorinta}},
  \bibinfo{author}{\bibfnamefont{J.~E.} \bibnamefont{Sipe}}, \bibnamefont{and}
  \bibinfo{author}{\bibfnamefont{M.}~\bibnamefont{Kauranen}},
  \bibinfo{journal}{Phys. Rev. B} \textbf{\bibinfo{volume}{80}},
  \bibinfo{pages}{233402} (\bibinfo{year}{2009}).

\bibitem[{\citenamefont{Liu et~al.}(2010)\citenamefont{Liu, Dai, Chin, and
  Zhang}}]{LiuNatPhot11}
\bibinfo{author}{\bibfnamefont{J.}~\bibnamefont{Liu}},
  \bibinfo{author}{\bibfnamefont{J.}~\bibnamefont{Dai}},
  \bibinfo{author}{\bibfnamefont{S.~L.} \bibnamefont{Chin}}, \bibnamefont{and}
  \bibinfo{author}{\bibfnamefont{X.-C.} \bibnamefont{Zhang}},
  \bibinfo{journal}{Nature Photonics} \textbf{\bibinfo{volume}{4}},
  \bibinfo{pages}{627} (\bibinfo{year}{2010}).

\bibitem[{\citenamefont{Dai et~al.}(2011)\citenamefont{Dai, Liu, and
  Zhang}}]{DaiIEEE11}
\bibinfo{author}{\bibfnamefont{J.}~\bibnamefont{Dai}},
  \bibinfo{author}{\bibfnamefont{J.}~\bibnamefont{Liu}}, \bibnamefont{and}
  \bibinfo{author}{\bibfnamefont{X.~C.} \bibnamefont{Zhang}},
  \bibinfo{journal}{IEEE Journal of Selected Topics in Quantum Electronics}
  \textbf{\bibinfo{volume}{17}}, \bibinfo{pages}{183} (\bibinfo{year}{2011}).

\bibitem[{\citenamefont{Bartels et~al.}(2009)\citenamefont{Bartels, Heinecke,
  and A~Diddams}}]{BartelsScience09}
\bibinfo{author}{\bibfnamefont{A.}~\bibnamefont{Bartels}},
  \bibinfo{author}{\bibfnamefont{D.}~\bibnamefont{Heinecke}}, \bibnamefont{and}
  \bibinfo{author}{\bibfnamefont{S.}~\bibnamefont{A~Diddams}},
  \bibinfo{journal}{Science} \textbf{\bibinfo{volume}{326}},
  \bibinfo{pages}{681} (\bibinfo{year}{2009}).

\bibitem[{\citenamefont{Rodrigo et~al.}(2013)\citenamefont{Rodrigo,
  Harutyunyan, and Novotny}}]{RodrigoPRL13}
\bibinfo{author}{\bibfnamefont{S.~G.} \bibnamefont{Rodrigo}},
  \bibinfo{author}{\bibfnamefont{H.}~\bibnamefont{Harutyunyan}},
  \bibnamefont{and} \bibinfo{author}{\bibfnamefont{L.}~\bibnamefont{Novotny}},
  \bibinfo{journal}{Phys. Rev. Lett.} \textbf{\bibinfo{volume}{110}},
  \bibinfo{pages}{177405} (\bibinfo{year}{2013}).

\bibitem[{\citenamefont{Grosse et~al.}(2012)\citenamefont{Grosse, Heckmann, and
  Woggon}}]{GrossePRL12}
\bibinfo{author}{\bibfnamefont{N.~B.} \bibnamefont{Grosse}},
  \bibinfo{author}{\bibfnamefont{J.}~\bibnamefont{Heckmann}}, \bibnamefont{and}
  \bibinfo{author}{\bibfnamefont{U.}~\bibnamefont{Woggon}},
  \bibinfo{journal}{Phys. Rev. Lett.} \textbf{\bibinfo{volume}{108}},
  \bibinfo{pages}{136802} (\bibinfo{year}{2012}).

\bibitem[{\citenamefont{Reddy et~al.}(2017)\citenamefont{Reddy, Chen,
  Fern\'{a}ndez-Dom\'{i}nguez, and Sivan}}]{NireekshanJOSAB17}
\bibinfo{author}{\bibfnamefont{K.~N.} \bibnamefont{Reddy}},
  \bibinfo{author}{\bibfnamefont{P.~Y.} \bibnamefont{Chen}},
  \bibinfo{author}{\bibfnamefont{A.~I.}
  \bibnamefont{Fern\'{a}ndez-Dom\'{i}nguez}}, \bibnamefont{and}
  \bibinfo{author}{\bibfnamefont{Y.}~\bibnamefont{Sivan}}, \bibinfo{journal}{J.
  Opt. Soc. Am. B} \textbf{\bibinfo{volume}{34}}, \bibinfo{pages}{1824}
  (\bibinfo{year}{2017}).

\bibitem[{\citenamefont{Agarwal and Jha}(1982)}]{AgarwalSolidStateCommun82}
\bibinfo{author}{\bibfnamefont{G.~S.} \bibnamefont{Agarwal}} \bibnamefont{and}
  \bibinfo{author}{\bibfnamefont{S.~S.} \bibnamefont{Jha}},
  \bibinfo{journal}{Solid State Commun.} \textbf{\bibinfo{volume}{41}},
  \bibinfo{pages}{499} (\bibinfo{year}{1982}).

\bibitem[{\citenamefont{Hua and Gersten}(1986)}]{HuaPRB86}
\bibinfo{author}{\bibfnamefont{X.~M.} \bibnamefont{Hua}} \bibnamefont{and}
  \bibinfo{author}{\bibfnamefont{J.~I.} \bibnamefont{Gersten}},
  \bibinfo{journal}{Phys. Rev. B} \textbf{\bibinfo{volume}{33}},
  \bibinfo{pages}{3756} (\bibinfo{year}{1986}).

\bibitem[{\citenamefont{\"Oestling et~al.}(1993)\citenamefont{\"Oestling,
  Stampfli, and Bennemann}}]{OestlingZPhysD93}
\bibinfo{author}{\bibfnamefont{D.}~\bibnamefont{\"Oestling}},
  \bibinfo{author}{\bibfnamefont{P.}~\bibnamefont{Stampfli}}, \bibnamefont{and}
  \bibinfo{author}{\bibfnamefont{K.~H.} \bibnamefont{Bennemann}},
  \bibinfo{journal}{Z. Phys. D: At., Mol. Clusters}
  \textbf{\bibinfo{volume}{28}}, \bibinfo{pages}{169} (\bibinfo{year}{1993}).

\bibitem[{\citenamefont{Dadap et~al.}(1999)\citenamefont{Dadap, Shan,
  Eisenthal, and Heinz}}]{DadapPRL99}
\bibinfo{author}{\bibfnamefont{J.~I.} \bibnamefont{Dadap}},
  \bibinfo{author}{\bibfnamefont{J.}~\bibnamefont{Shan}},
  \bibinfo{author}{\bibfnamefont{K.~B.} \bibnamefont{Eisenthal}},
  \bibnamefont{and} \bibinfo{author}{\bibfnamefont{T.~F.} \bibnamefont{Heinz}},
  \bibinfo{journal}{Phys. Rev. Lett.} \textbf{\bibinfo{volume}{83}},
  \bibinfo{pages}{4045} (\bibinfo{year}{1999}).

\bibitem[{\citenamefont{Butet et~al.}(2010)\citenamefont{Butet, Duboisset,
  Bachelier, Russier-Antoine, Benichou, Jonin, and Brevet}}]{ButetNanoLett10}
\bibinfo{author}{\bibfnamefont{J.}~\bibnamefont{Butet}},
  \bibinfo{author}{\bibfnamefont{J.}~\bibnamefont{Duboisset}},
  \bibinfo{author}{\bibfnamefont{G.}~\bibnamefont{Bachelier}},
  \bibinfo{author}{\bibfnamefont{I.}~\bibnamefont{Russier-Antoine}},
  \bibinfo{author}{\bibfnamefont{E.}~\bibnamefont{Benichou}},
  \bibinfo{author}{\bibfnamefont{C.}~\bibnamefont{Jonin}}, \bibnamefont{and}
  \bibinfo{author}{\bibfnamefont{P.-F.} \bibnamefont{Brevet}},
  \bibinfo{journal}{Nano Letters} \textbf{\bibinfo{volume}{10}},
  \bibinfo{pages}{1717} (\bibinfo{year}{2010}).

\bibitem[{\citenamefont{McMahon et~al.}(2006)\citenamefont{McMahon, Lopez,
  Haglund, Ray, and Bunton}}]{McMahonPRB06}
\bibinfo{author}{\bibfnamefont{M.~D.} \bibnamefont{McMahon}},
  \bibinfo{author}{\bibfnamefont{R.}~\bibnamefont{Lopez}},
  \bibinfo{author}{\bibfnamefont{R.~F.} \bibnamefont{Haglund}},
  \bibinfo{author}{\bibfnamefont{E.~A.} \bibnamefont{Ray}}, \bibnamefont{and}
  \bibinfo{author}{\bibfnamefont{P.~H.} \bibnamefont{Bunton}},
  \bibinfo{journal}{Phys. Rev. B} \textbf{\bibinfo{volume}{73}},
  \bibinfo{pages}{041401} (\bibinfo{year}{2006}).

\bibitem[{\citenamefont{van Nieuwstadt et~al.}(2006)\citenamefont{van
  Nieuwstadt, Sandtke, Harmsen, Segerink, Prangsma, Enoch, and
  Kuipers}}]{NieuwstadtPRL06}
\bibinfo{author}{\bibfnamefont{J.~A.~H.} \bibnamefont{van Nieuwstadt}},
  \bibinfo{author}{\bibfnamefont{M.}~\bibnamefont{Sandtke}},
  \bibinfo{author}{\bibfnamefont{R.~H.} \bibnamefont{Harmsen}},
  \bibinfo{author}{\bibfnamefont{F.~B.} \bibnamefont{Segerink}},
  \bibinfo{author}{\bibfnamefont{J.~C.} \bibnamefont{Prangsma}},
  \bibinfo{author}{\bibfnamefont{S.}~\bibnamefont{Enoch}}, \bibnamefont{and}
  \bibinfo{author}{\bibfnamefont{L.}~\bibnamefont{Kuipers}},
  \bibinfo{journal}{Phys. Rev. Lett.} \textbf{\bibinfo{volume}{97}},
  \bibinfo{pages}{146102} (\bibinfo{year}{2006}).

\bibitem[{\citenamefont{Manjare et~al.}(2017)\citenamefont{Manjare, Wang,
  Rodrigo, and Harutyunyan}}]{ManjareAPL17}
\bibinfo{author}{\bibfnamefont{M.}~\bibnamefont{Manjare}},
  \bibinfo{author}{\bibfnamefont{F.}~\bibnamefont{Wang}},
  \bibinfo{author}{\bibfnamefont{S.~G.} \bibnamefont{Rodrigo}},
  \bibnamefont{and}
  \bibinfo{author}{\bibfnamefont{H.}~\bibnamefont{Harutyunyan}},
  \bibinfo{journal}{Applied Physics Letters} \textbf{\bibinfo{volume}{111}},
  \bibinfo{pages}{221106} (\bibinfo{year}{2017}).

\bibitem[{\citenamefont{Economou}(1969)}]{EconomouPhysRev69}
\bibinfo{author}{\bibfnamefont{E.~N.} \bibnamefont{Economou}},
  \bibinfo{journal}{Phys. Rev.} \textbf{\bibinfo{volume}{182}},
  \bibinfo{pages}{539} (\bibinfo{year}{1969}).

\end{thebibliography}

\end{document}